\documentstyle[prb,aps,psfig]{revtex}
\newcommand \be{\begin{eqnarray}}
\newcommand \ee{\end{eqnarray}}
\begin{document}
\twocolumn[\hsize\textwidth\columnwidth\hsize
           \csname @twocolumnfalse\endcsname
\title{Ginzburg-Landau theory of superconducting surfaces under electric fields } 
\author{P. Lipavsk\'y$^{1,2}$, K. Morawetz$^{3,4}$, J. Kol\'a\v cek$^2$, 
T.~J.~Yang$^5$}
\address{
$^1$ Faculty of Mathematics and Physics, Charles University, 
Ke Karlovu 3, 12116 Prague 2, Czech Republic\\
$^2$Institute of Physics, Academy of Sciences, 
Cukrovarnick\'a 10, 16253 Prague 6, Czech Republic\\
$^3$Institute of Physics, Chemnitz University of Technology, 
09107 Chemnitz, Germany\\
$^4$Max-Planck-Institute for the Physics of Complex
Systems, Noethnitzer Str. 38, 01187 Dresden, Germany\\
$^5$Department of Electrophysics, National Chiao Tung University,
Hsinchu 300, Taiwan}
\maketitle
\begin{abstract}
A boundary condition for the Ginzburg-Landau wave function at
surfaces biased by a strong electric field is derived 
within the de Gennes approach. This condition provides a simple theory
of the field effect on the critical temperature of superconducting layers.
\end{abstract}
    \vskip2pc]

The critical temperature of a thin superconducting layer 
is increased or lowered by an electric field applied 
perpendicular to the layer.\cite{GS60,XDWKLV92,FreyMannhart95,ATM03,MGT03}
Similarly to the conductivity of inverse layers in semiconductors, 
superconductivity of thin metallic layers can thus be controlled 
by a gate voltage, which makes these structures attractive for 
applications. 

In this paper we show that the phase transition in a 
thin metallic layer is conveniently described by the 
Ginzburg-Landau (GL) theory, where the electric field 
$E$ enters the GL boundary condition as
\begin{equation}
\left.{\nabla\psi\over\psi}\right|_0=
\left.{\nabla\Delta\over\Delta}\right|_0=
{1\over b}={1\over b_0}+{E\over U_{\rm s}}.
\label{e1}
\end{equation}
Briefly, the logarithmic derivative of the GL function  
$\psi$ or the gap function $\Delta$ at the surface is 
a sum of the zero-field part $1/b_0$ and the field induced 
correction $E/U_{\rm s}$.

The zero-field part has been derived by de Gennes\cite{G66}
from the BCS theory. A typical value $b_0\sim 1$~cm is large 
on the scale of the GL coherence length, therefore this 
contribution is usually neglected. This approximation, 
$1/b_0\approx 0$, corresponds to the original GL condition 
\mbox{$\nabla\psi=0$}. 

Here we employ the de Gennes approach to derive the field induced 
correction $E/U_{\rm s}$.  The correction becomes important for 
the above mentioned experiments, where fields of the order of 
10$^7$ V/cm are realized. Small electric fields appearing
e.g. in Josephson junctions do not require such corrections.

We start from the condition
\begin{equation}
{1\over b}={1\over\xi^2(0)}
{1\over N_0V}
\int\limits_{-\infty}^\infty dx{\Delta(x)\over\Delta_0}
\left[1-{N(x)\over N_0}\right]
\label{e2}
\end{equation}
derived by de Gennes (Eq. (7-62) in Ref.~\onlinecite{G66}).
Here $N_0$ is the density of states of a bulk material, $V$ is 
the BCS interaction, and $N(x)$ is the local density of
states at position $x$. The actual gap function $\Delta(x)$ has 
a non-trivial profile close to the surface at $x=0$, but it has
only slow variation at distances exceeding the BCS coherence 
length $\xi_0=0.18\,\hbar v_{\rm F}/k_{\rm B}T_{\rm c}$. For 
$x\sim\xi_0$ it is crudely linear $\Delta(x)\approx\Delta_0\left(
1+x/b\right)$, so that $\Delta_0$ is not the true surface
value but the extrapolation of the gap function towards the 
surface. In Eq. (\ref{e2}) we have used the GL coherence
length at zero temperature $\xi(0)=0.74\,\xi_0$ for pure metals.

In measurements of the field effect on the transition 
temperature, the zero-field term $b_0$ is included in the reference 
zero-bias transition temperature. Accordingly, we can assume 
a model of the crystal for which $1/b_0=0$. The simplest model 
of this kind is a semi-infinite jellium, where for zero field the 
density of states is step-like, $N(x)=N_0$ for $x>0$ and $N(x)=0$ 
elsewhere. Using that the gap 
function is restricted to the crystal, $\Delta(x)=0$ for $x<0$, 
one can check that from (\ref{e2}) follows $1/b_0=0$.

Now we include the electric field. According to the Anderson 
theorem \cite{Anderson59}, the electric field does not change the 
thermodynamical properties directly but only via the density 
of states. The change of the density of states is also indirect.
The penetrating electric field induces a deviation $\delta n$ of 
the electron density. The density deviation changes the Fermi 
momentum. Since the density of states depends on the Fermi 
momentum, its value becomes modified. We express this complicated 
indirect effect approximatively via a local linear expansion
\begin{equation}
N(x)=N_0+{\partial N_0\over\partial n}\delta n(x).
\label{e3}
\end{equation}
The de Gennes condition (\ref{e2}) then reads
\begin{equation}
{E\over U_{\rm s}}=-{1\over\xi^2(0)}{1\over N_0^2V}
{\partial N_0\over\partial n}
\int\limits_0^\infty dx{\Delta(x)\over\Delta_0}
\delta n(x).
\label{e4}
\end{equation}

The actual space profile of $\delta n$ in superconductors is 
unknown. In fact, some of recent measurements suggests that the 
electric field penetrates deep into superconductors.\cite{Tao03}
Interpretation of these observations is not yet settled, 
therefore we prefer to assume that the screening 
in superconductors is similar to the screening in normal metals 
so that $\delta n$ is non-zero only on the scale of the Thomas-Fermi 
screening length. The typical Thomas-Fermi length is less then 
one \AA ngstr\"om, while the gap function varies on a scale typical 
to the BCS kernel $\sim\xi_0$. Accordingly, in the integral 
(\ref{e4}) we can take $\Delta(x)\approx\Delta(0)$ and obtain
\begin{equation}
{1\over U_{\rm s}}={1\over\xi^2(0)}{1\over N_0^2V}
{\partial N_0\over\partial n}{\Delta(0)\over\Delta_0}
{\epsilon_0\over e}.
\label{e5}
\end{equation}
In this rearrangement we have used the surface charge 
determined by the applied field $\epsilon_0 E=-e\int_0^\infty 
dx~\delta n(x)$.

The effective potential $U_{\rm s}$ given by (\ref{e5}) depends 
on bulk material parameters $\xi_0$, $N_0V$ and $\partial N_0/
\partial n$, and on the ratio of the gap at the surface to the 
bulk value 
\begin{equation}
\eta={\Delta(0)\over\Delta_0}. 
\label{e5a}
\end{equation}
According to de Gennes estimates\cite{G66}, the surface ratio 
$\eta$ is of the order of unity. A heuristic derivation of the 
field-effect from the GL equation cannot cover this factor.\cite{Lee96}

It is advantageous to express the effective potential $U_{\rm s}$
via the usual parameters of the GL theory. First we employ the
BCS relation for the critical temperature $k_{\rm B}T_{\rm c}=
1.14~\hbar\omega_{\rm D}~\exp\left(-1/N_0V\right)$. The critical 
temperature depends on the density of electrons. Comparing alloys 
with different impurity doping, it has been deduced that the 
dominant density dependence enters the critical temperature via 
the density of states.\cite{Varm76} We can thus assume 
$\partial\omega_{\rm D}/\partial n\approx 0$ and $\partial V/
\partial n\approx 0$ with the help of which we express the 
derivative of the density of states via the logarithmic 
derivative of the critical temperature. Formula (\ref{e5}) 
then simplifies to
\begin{equation}
{1\over U_{\rm s}}=\eta~\kappa^2~
{\partial\ln T_{\rm c}\over\partial \ln n}
{e\over mc^2}.
\label{e6}
\end{equation}
Here we have expressed the electron density via the London
penetration depth $\lambda^2(0)=m/(\mu_0 ne^2)$. Its ratio
to the GL coherence length defines the GL parameter
$\kappa=\lambda(0)/\xi(0)$.

Let us estimate the effective potential $U_{\rm s}$ for
niobium. The charge carriers are electrons, $e=-|e|$,
with the mass close to the electron rest mass,
$m\approx 1.2\,m_{\rm e}$. The GL parameter is on the edge 
of the type-I and II materials, $\kappa=0.78$, and the logarithmic
derivative is of moderate amplitude, $\partial\ln T_{\rm c}/
\partial \ln n=0.75$ (see Ref.~\onlinecite{LKMB02}). Taking 
$\eta\approx 1$ one finds, 
$U_{\rm s}=-1.3~10^6$~V. As one can see, a large field 
$E\sim 10^6$~V/cm is necessary to create 
a field-induced correction at least comparable to the 
commonly neglected zero-field value $1/b_0\sim 1/$cm.

The effective potential (\ref{e6}) is the major result of
this paper. Now we use it in the boundary condition (\ref{e1}) 
to evaluate the transition temperature $T^*$ of a biased 
layer of a finite thickness $L$. General steps of our analysis 
parallel the theory of the Little-Parks effect \cite{Tinkham}. 
It is also in a close analogy to the theory of surface 
superconductivity in short coherence length materials 
\cite{Chen94}.

Let us assume that the electric field is applied only to the 
left surface at $x=0$, while the right surface at $x=L$ is 
free of the field. We take $1/b_0=0$  for simplicity, so that 
we use the boundary conditions
\begin{eqnarray}
\left.{\nabla\psi\over\psi}\right|_0&=&{E\over U_{\rm s}}~,
\label{e7}\\
\left.{\nabla\psi\over\psi}\right|_L&=&0~.
\label{e8}
\end{eqnarray}

The GL function is given by the dimension-less GL equation 
$\xi^2(T)\nabla^2\psi+\psi-|\psi|^2\psi=0$, see Ref~\onlinecite{Tinkham}.
The transition point is characterized by an infinitesimally small 
GL function, $|\psi|^2\to 0$. At the transition 
temperature $T^*$, the non-linear term in the GL equation 
thus vanishes $\xi^2(T^*)\nabla^2\psi+\psi=0$. This equation
is solved by $\psi(x)\propto\cos\left[(x-L)/\xi(T^*)\right]$,
which satisfies the right boundary condition (\ref{e8}) 
while the left boundary condition (\ref{e7}) demands
\begin{equation}
{L\over\xi(T^*)}\tan\left({L\over\xi(T^*)}\right)
={EL\over U_{\rm s}}.
\label{e13}
\end{equation}

\begin{figure}
\psfig{file=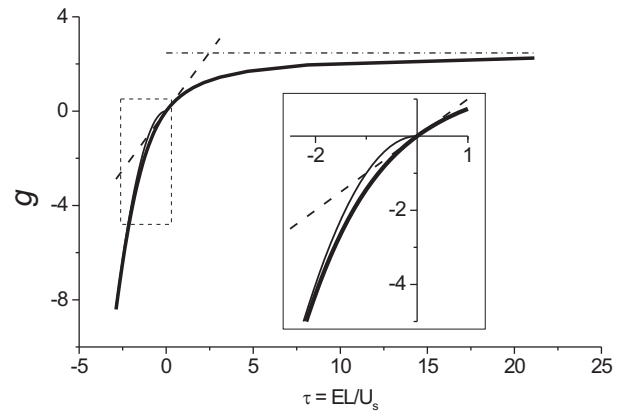,width=9cm}
\caption{The dimensionless shift of the transition
temperature due to the electric field as given by
Eq.~(\protect{\ref{e14}}). The exact 
solution of $\sqrt{g}\tan\sqrt{g}=\tau$ (thick full line), the 
linear approximation $g\approx\tau$ for thin layers 
(tangential dashed line), the constant approximation 
$g\approx\pi^2/4$ for large suppressive fields (dot-dashed 
line), and the parabolic approximation $g\approx-\tau^2$
for large supportive fields (thin full line). The insert
shows a detail of the parabolic approximation.
\label{fig1}}
\end{figure}

When the superconductor has a coherence length $\xi$ which
satisfies the condition (\ref{e13}), the non-zero GL wave
function nucleates and the system undergoes a transition
to the superconducting state. Since the coherence length is 
a function of temperature,
\begin{equation}
\xi(T)={\xi(0)\over\sqrt{1-{T\over T_{\rm c}}}},
\label{e10}
\end{equation}
one can find from (\ref{e13}) and (\ref{e10}) the transition
temperature $T^*$. It reads
\begin{equation}
T^*=T_{\rm c}-T_{\rm c}{\xi^2(0)\over L^2}
g\left({EL\over U_{\rm s}}\right),
\label{e14}
\end{equation}
where the function $g(\tau)$ is a root of $\sqrt{g}\,
\tan\sqrt{g}=\tau$. The function $g$ is plotted in 
figure~\ref{fig1}.

Although equation (\ref{e14}) is simple by itself, we find it
useful to discuss its asymptotic solutions. Let us start with
the experimentally most important limit. The effect of the 
electric field on the transition temperature is rather small 
and it is most conveniently observed on very thin layers. In 
this case  $|EL|\ll|U_s|$ and one can use the linear approximation
$g(\tau)\approx\tau$ shown as the dashed line in Fig~\ref{fig1}.
Within linear approximation the transition temperature 
(\ref{e10}) simplifies to
\begin{equation}
T^*=T_{\rm c}-\eta~{\partial T_{\rm c}\over\partial n}
{\epsilon_0E\over eL}.
\label{e16}
\end{equation}

Formula (\ref{e16}) shows that surface ratio (\ref{e5a})
reduces field effect on the transition temperature $T^*$.
One can compare (\ref{e16}) with a simple estimate that 
assumes that the induced charge is homogeneously distributed 
across the layer, $e\delta n(x)\approx-\epsilon_0E/L$. 
A bulk critical temperature modified by the excess charge
is then interpreted as the transition temperature, 
$T^*=T_{\rm c}+\partial T_{\rm c}/\partial n\times\delta n$. 
Apparently, the simple estimate does not include the
surface ratio $\eta$.

Based on the simple estimate one is tempted to say that 
$\partial T_{\rm c}/\partial n$ can be uniquely determined 
from experimental data on $T^*$. The present theory shows, 
however, that surface ratio obscures the observed value.

Among high-$T_{\rm c}$ superconductors there are many 
materials of large $\kappa$. From equation (\ref{e6}) one can 
see that these materials have much lower effective potential 
$U_{\rm s}$, therefore they reveal a much stronger field effect on 
the transition temperature. With these materials it is possible to 
achieve the opposite limit -- the regime of thick layers
$|E L|\gg|{U}_{\rm s}|$. 

A measurement in the regime of thick layers has been already performed 
by Matijasevic {\em et al}.\cite{Matijasevic94}  We will use parameters
of their sample  to evaluate effects expected from formula
(\ref{e14}). The sample Sm$_{0.7}$Ca$_{0.3}$Cu$_3$O$_y$ is overdoped 
with $T_{\rm c}$ reduced 
to 50~K. The carriers are holes, $e=|e|$, of the mass
$m=3.46\,m_{\rm e}$ and the density $n=5.75~10^{21}$/cm$^3$.
Based on the authors claim that $T_{\rm c}$ in a monolayer would 
increase by 10~K at the imposed voltage, one can deduce 
$\partial\ln T_{\rm c}/\partial\ln n=-3.12$. With a typical GL 
parameter $\kappa=100$ one obtains the GL coherence length 
$\xi(0)=1.3$~nm and an effective potential $U_{\rm s}=-56.6$~V.
Note that this is by four orders of magnitude smaller than the 
niobium value. The applied
field is enhanced by a large dielectric function to an effective 
value $E=7.8~10^7$~V/cm. For a sample width of $L=50$~nm and 
$\eta\approx 1$ we find $|EL/U_{\rm s}|=
6.9$, which confirms that these measurements are in the thick 
layer limit.

In the thick layer limit one has to distinguish whether the 
electric field supports or depresses the transition temperature. 
Let us first discuss the depression which appears for 
$\partial T_{\rm c}/\partial n\times\epsilon_0E/e>0$. In this 
case $EL/U_{\rm s}>0$ and the function $g$ approaches the 
constant asymptotic value $g\to\pi^2/4=2.47$ shown as the dash-dotted 
line in Fig.~\ref{fig1}. Since $g<\pi^2/4$, relation (\ref{e14}) yields 
the upper estimate of the depression of the transition temperature 
\begin{equation}
T_{\rm c}-T^*< T_{\rm c}{\pi^2\xi^2(0)\over 4L^2}.
\label{e19}
\end{equation}
One can see that the maximal depression is
limited by the layer thickness. Within the adopted approximations
the actual value of the electrostatic field does not matter once
the asymptotic regime is reached. For parameters of Ref.~\onlinecite{Matijasevic94}
one finds from formula (\ref{e19}) the lower estimate  $T_{\rm c}-T^*<0.08$~K. 
The formula (\ref{e14}) gives a slightly smaller value, as one can guess from 
Fig.~\ref{fig1}. For $|EL/U_{\rm s}|=6.9$ the dimensionless shift is 
$g(6.9)=1.89$, i.e., $T^*-T_{\rm c}=-0.064$~K. Matijasevic 
{\em et al}\cite{Matijasevic94} reported no suppression of the superconductivity. 
Our estimate shows that the suppression is below the sensitivity of their 
method.

A different situation is met if the direction of the electric field
is reversed. The electric field then supports the superconductivity 
since $\partial T_{\rm c}/\partial n\times\epsilon_0E/e<0$. 
In this case $EL/U_{\rm s}<0$ and the function $g$ approaches
its quadratic asymptotics, $g\to-\tau^2$ shown as the thin full line 
in Fig.~\ref{fig1}. Since $g=L^2/\xi^2(T^*)$, the coherence length 
$\xi(T^*)$ is imaginary giving the GL function exponentially 
decaying from the biased surface. In this limit,
\begin{equation}
T^*\to T_{\rm c}+T_{\rm c}{E^2\xi^2(0)\over U_{\rm s}^2},
\label{e22}
\end{equation}
the critical temperature does not depend on the width of sample and
increases quadratically with the electric field. For parameters of
Ref.~\onlinecite{Matijasevic94} we obtain $T^*-T_{\rm c}=1.6$~K in a 
reasonable agreement with the reported shift by 1~K. 

The increased critical temperature (\ref{e22}) is independent of the 
layer width $L$. This shows that in this limit the superconductivity 
above $T_{\rm c}$ is stimulated in the same way as the 
superconductivity on the surface of an infinite sample predicted by 
Shapiro~\cite{Shapiro85}. Formula (\ref{e22}) differs from Shapiro's
formula (9) only by a factor due to the impurity limited coherence
length assumed in Ref.~\onlinecite{Shapiro85}.

We note that the present discussion does not account for the 
charge reservoirs typical to layered CuO materials. For a 
microscopic study devoted to these materials see 
Ref.~\onlinecite{KonsinSorkin01}. We also do not assume an eventual
%KonsinSorkin01a
effect of the electric field on the chemical composition,
e.g., due to oxygen motion as proposed in 
Refs.~\onlinecite{CVG93,Grig96}.

In summary, using the de Gennes approach we have derived the GL 
boundary condition for a superconducting surface exposed to the 
electric field. This boundary condition allows one to conveniently 
evaluate the field effect on surface sensitive phenomena from the
GL theory. Its implementation is demonstrated for the field effect 
on the superconducting phase transition in metallic layers. Our
approach recovers known features, in particular, that for thin
layers the transition temperature can be linearly enhanced or 
suppressed depending on the orientation of the applied field. We
have found, however, that compared to former theories the linear 
coefficient is modified by the value of the gap at the surface. 
In the limit of thick layers we obtain a field induced surface 
superconductivity with the shift of the critical temperature 
depending on the square of the electric field. We also obtain
the upper limit on the suppression of the critical temperature
being independent of the field and inversely proportional to the 
square of the layer width. These features agree with
the experimental data.

\acknowledgements
Authors are grateful to P. Martinoli who brought this problem
to their attention. This work was supported by 
GA\v{C}R 202/04/0585, 202/05/0173, GAAV A1010312 grants, by 
National Science Council of  Republic of
China, Taiwan, with grant Nsc 94-2112-M-009-001,
and by DAAD project D/03/44436. The 
European ESF program AQDJJ is also acknowledged. 

%\bibliography{kmsr,kmsr1,kmsr2,kmsr3,kmsr4,kmsr5,kmsr6,kmsr7,delay2,spin,sem1,sem2,micha,genn}

\begin{thebibliography}{10}

\bibitem{GS60}
R.~E. Glover and M.~D. Sherrill, Phys. Rev. Lett. {\bf 5},  248  (1960).

\bibitem{XDWKLV92}
X.~X. Xi {\it et~al.}, Phys. Rev. Lett. {\bf 68},  1240  (1992).

\bibitem{FreyMannhart95}
T. Frey, J. Mannhart, J.~G. Bednorz, and E.~J. Williams, Phys. Rev. B {\bf 51},
   3257  (1995).

\bibitem{ATM03}
C.~H. Ahn, J.~M. Triscone, and J. Mannhart, Nature {\bf 424},  1015  (2003).

\bibitem{MGT03}
D. Matthey, S. Gariglio, and J.~M. Triscone, Appl. Phys. Lett. {\bf 83},  3758
  (2003).

\bibitem{G66}
P.~G. de~Gennes, {\em Superconductivity of Metals and Alloys} (Benjamin, New
  York, 1966), Chap.~VII.3.

\bibitem{Anderson59}
P.~W. Anderson, J. Phys. Chem. Solids {\bf 11},  26  (1959).

\bibitem{Tao03}
R. Tao, X. Xu, and E. Amr, Phys. Rev. B {\bf 68},  144505  (2003).

\bibitem{Lee96}
W.~D. Lee, J.~L. Chen, T.~J. Yang, and B.-S. Chiou, Physica C {\bf 261},  167
  (1996).

\bibitem{Varm76}
C.~M. Varma and R.~C. Dynes,  in {\em Superconductivity in d- and f-band
  Metals}, edited by D.~H. Douglass (Plenum Press, New York, 1976).

\bibitem{LKMB02}
P. Lipavsk{\'y}, J. Kolacek, K. Morawetz, and E.~H. Brandt, Phys. Rev. B {\bf
  66},  134525  (2002).

\bibitem{Tinkham}
M. Tinkham, {\em Introduction to Superconductivity} (McGraw Hill, New York,
  1966), Chap.~4.

\bibitem{Chen94}
J.~L. Chen and T.~J. Yang, Physica C {\bf 231},  91  (1994).

\bibitem{Matijasevic94}
V.~C. Matijasevic {\it et~al.}, Physica C {\bf 235},  2097  (1994).

\bibitem{Shapiro85}
B.~Y. Shapiro, Solid State Commun. {\bf 53},  673  (1985).

\bibitem{KonsinSorkin01}
P. Konsin and B. Sorkin, Phys. Rev. B {\bf 58},  5795  (1998).

\bibitem{CVG93}
N. Chandrasekhar, O.~T. Valls, and A.~M. Goldman, Phys. Rev. Lett. {\bf 71},
  1079  (1993).

\bibitem{Grig96}
G. Grigelionis, E.~E. Tornau, and A. Rosengren, Phys. Rev. B {\bf 53},  425
  (1996).

\end{thebibliography}
%\bibliographystyle{prsty}
%\end{document}

\end{document}